\documentclass{ws-procs9x6}

\begin{document}

\title{Deformations of plane algebraic curves and integrable
systems of hydrodynamic type}

\author{YUJI KODAMA\footnote{\uppercase{YK} was partially
supported by {\uppercase{NSF}} grant {\uppercase{DMS}}0071523.}}

\address{Department of Mathematics, \\
Ohio State University, \\
Columbus, OH 43210, USA\\
E-mail: kodama@math.ohio-state.edu}

\author{BORIS G. KONOPELCHENKO\footnote{\uppercase{BGK} was partially
supported by grant {\uppercase{COFIN}} 2000 '{\uppercase{S}}intesi'.}}

\address{Dipartimento di Fisica, \\
Universita di Lecce and Sezione INFN, \\
73100 Lecce, Italy\\
E-mail: konopel@le.infn.it}

\maketitle

\abstracts{
We describe our recent work on deformations of hyperelliptic
curves by means of integrable hierarchy of hydrodynamic type, and
discuss a further extension to the cases of non-hyperelliptic
curves. }

\section{Deformation of plane algebraic curves}
Let $\mathcal C$ be a plane algebraic curve on ${\mathbb C}^2=\{(k,p)\}$:
\begin{equation}
\label{curve}
{\mathcal C}:~~ p^N=\sum_{j=1}^{N}u_j(k)p^{N-j}, \quad u(k)\in {\mathbb C}[k].
\end{equation}
Then consider a deformation of the curve ${\mathcal C}$ as the following
equation of conservation law;
\begin{equation}
\label{deformation}
\frac{\partial p}{\partial t}=\frac{\partial}{\partial x}Q, \quad 
{\rm with} \quad Q\in \frac{{\mathbb C}[k,p]}{\mathcal 
C}=\bigoplus_{k=1}^N
{\mathbb C}[k]\cdot p^{N-k},
\end{equation}
where $x,t$ are the deformation parameters.
The equation (\ref{deformation}) describes a (quasi-linear)
system of equations of hydrodynamic type for $u_j=u_j(x,t;k)$ in the 
curve $\mathcal C$.
Then from a direct computation we find:
\begin{lemma}
\label{Ndef}
Let $Q$ be given by
\[
Q=\sum_{j=1}^{N}\alpha_j(k)p^{N-j}, \quad \alpha_j(k)\in {\mathbb C}[k].
\]
Then the functions $u_j$ in the curve ${\mathcal C}$ satisfy
\[
\frac{\partial u_l}{\partial t}=\sum_{j=1}^N 
N\gamma_l^{(j)}\frac{\partial \alpha_j}{\partial x} + 
\sum_{i=1}^{N}\sum_{j=1}^N(N-j)\gamma_l^{(i+j)}
\left(\alpha_j\frac{\partial u_i}{\partial x}-u_j\frac{\partial 
\alpha_i}{\partial x}\right),
\]
where $\gamma_l^{(j)}$ are defined by
\[
p^{2N-j-1}=\sum_{l=1}^N\gamma_l^{(j)}p^{N-k}, \quad 1\le j\le 2N-1.
\]
\end{lemma}
The last equation leads to the recursion relation
for $\gamma_l^{(i)}$ for $l=1,\cdots,N$ with $\gamma_{N+1}^{(i)}=0$,
\[
\gamma_l^{(i-1)}=
\left\{\begin{array}{llll}
  u_l\gamma_1^{(i)}+\gamma_{l+1}^{(i)}, \quad &{\rm for}
\quad &1\le i \le N-1\\
\delta_{l,i-N+1}, \quad &{\rm for}\quad & N\le i\le 2N-1
\end{array}
\right.
\]
which can be used to find the explicit form of $\gamma_l^{(i)}$'s
in terms of $u_j$'s.

A hierarchy associated with the system (\ref{deformation}) may be obtained by
choosing proper $u_j(k),~\alpha_j(k)\in {\mathbb C}[k]$ for $j=1,\cdots,N$,
so that the systems for $n=1,2,\cdots$,
\[
\frac{\partial p}{\partial t_n}=\frac{\partial }{\partial x}Q^{(n)},
\quad {\rm with }\quad Q^{(n)}=\sum_{j=1}^N\alpha_j^{(n)}(k)p^{N-j},
\]
are compatible, i.e. $\partial^2p/\partial t_n\partial t_m=
\partial^2p/\partial t_m\partial t_n$. (In this paper we choose 
$Q^{(0)}=p$, that is,
${\partial p}/{\partial t_0}={\partial Q^{(0)}}/{\partial x}$,
and identify $t_0=x$.) A general scheme to construct such
$Q^{(n)}$ has not been found. It is however quite interesting to classify
the curves which allow integrable deformations of this type.
In this report, we give several examples
of the hierarchy associated with the deformations of the curve ${\mathcal C}$
in (\ref{curve}).

\begin{remark}
\label{LGrem}
The curve (\ref{curve}) contains several known examples of the hierarchy,
which are obtained by the reductions of the dispersionless KP (and mKP) hierarchy
(see also Remark \ref{Nrem}).
The examples contain a model of 
  the Landau-Ginzburg theory with rational potentials
\cite{aoyama:96}.
This is simply obtained by the following choices
of polynomials for $u_i$'s in (\ref{curve}):
Fix an index of $u_i$'s, say $n$, and set
\begin{equation}
\label{dmkp}
u_n=k^n-v_n, \quad {\rm and}\quad u_i=-v_i,
\quad {\rm for}\quad i\ne n,
\end{equation}
where $v_i$'s do not depend on $k$.
Then we have
\begin{equation}
\label{LG}
\displaystyle{k^n=p^n+v_1p^{n-1}+\cdots + v_n +\frac{v_{n+1}}{p}+\cdots
+\frac{v_N}{p^{N-n}}.}
\end{equation}
which gives a rational Landau-Ginzburg potential\cite{aoyama:96}.
An interesting question is how one can extend
the ring ${\mathbb C}[k,p]/{\mathcal C}$ to a rational ring. Also one should
include a log-term to define a complete set of primary fields.

There is another interesting example, which is obtained by
the dipersionless limit of the vector nonlinear Schr\"odinger 
equation\cite{zakharov:80}
(Zhakharov reduction\cite{kodama:89}),
\[
k=p+\sum_{j=1}^{N-1}\frac{\rho_j}{p-v_j}.
\]
The hierarchy associated to this curve provides a model of the topological
field theory with $N-1$ punctures.
\end{remark}

\begin{remark}
The system (\ref{deformation}) has an infinite number of conserved
densities which are explicitely obtained by expressing $p$ in the
Laurent series of $k$.
\end{remark}

\section{Deformations of hyperelliptic curves $(N=2)$}
The curve ${\mathcal C}$ with $N=2$ corresponds to hyperelliptic curve
of genus determined by the degrees of polynomials $u_j(k),~ j=1,2$;
\[
p^2=u_1(k)p+u_2(k), \quad u_1, u_2\in {\mathbb C}[k].
\]
The deformation equation (\ref{deformation}) then gives
\begin{equation}
\label{N2system}
\left\{\begin{array}{llll}
&\displaystyle{\frac{\partial u_1}{\partial t} = \frac {\partial}{\partial x}
\left(u_1\alpha_1+2\alpha_2\right)} \\
&{}\\
&\displaystyle{\frac{\partial u_2}{\partial x} = 2u_2\frac{\partial 
\alpha_1}{\partial x}+
\alpha_1\frac{\partial u_2}{\partial t}-u_1\frac{\partial 
\alpha_2}{\partial x}.}
\end{array}
\right.
\end{equation}
As a particular case with $u_1=k-v_1,~u_2=-v_2$ (see Remark 
\ref{LGrem}), we have
\[
k=p+v_0+\frac{v_1}{p},
\]
which corresponds to the curve of the dispersioness Toda equation
discussed in \cite{kodama:90}. This curve is also related to the
dispersionless AKNS system for the variable $p'=p+v_0$.

We also note that the variable $u_1(k)$ can be eliminated by the
(gauge) transformation, $p\to p+u_1/2$, and then from
(\ref{N2system}) one can take $Q$ in the form,
\[
Q=\alpha_1(k)p, \quad \quad ({\rm i.e.}~~\alpha_2=0).
\]
Thus we have the following two examples for the case $N=2$,
where the polynomial $u_2(k)$ is either odd or even.

\subsection{The ${\rm BH}_m$ hierarchy}
The ${\rm BH}_m$ hierarchy\cite{kodama:02} is defined on the
singular sectors of the Burgers-Hopf (BH) hiearchy, where the solution of
the hiearachy has a shock type singularity.
Here the number $m$ is the genus of the curve ${\mathcal C}$
given by
\begin{equation}
\label{BHcurve}
y^2=u_2(k)=k^{2m+1}+\sum_{j=0}^{2m}v_jk^{2m-j}.
\end{equation}
In particular, the case with $m=0$ gives the BH hierarchy
(or the dispersionless KdV hierarchy), i.e. with $u_2=k+v_0$,
\[
p^2=k+v, \quad {\rm and} \quad Q^{(n)}=\alpha_1^{(n)}(k)p=[k^{n+\frac{1}{2}}]_{+p},
\]
where $[k^{a}]_{+p}$ represents the polynomial part of $k^a$ in $p$.
The BH hierarchy in terms of the variable $v$ is given by
\[
\frac{\partial v}{\partial t_n}=c_nv^n\frac{\partial v}{\partial x}, \quad
{\rm with}\quad c_n=(-1)^n\frac{(2n+1)!!}{2^nn!},\quad n=0,1,2,\cdots.
\]

In the case with $m\ne 0$, we have\cite{kodama:02}:
\begin{proposition}
\label{BHm}
The following system associated with the curve (\ref{BHcurve})
forms an integrable hierachy of a deformation of the curve;
\[
\frac{\partial p}{\partial t_n}=\frac{\partial}{\partial x}
Q^{(n)}, \quad {\rm with} \quad Q^{(n)}=\alpha_1^{(n)}p=
\left[\frac{k^{m+n+\frac{1}{2}}}{p}\right]_+p,
\]
where $[\cdot]_+$ is the polynomial part of $k$.
\end{proposition}
The form $Q^{(n)}$ can be obtained as follows: First assume $\alpha_1^{(n)}(k)$
be a monic polynomial of degree $n$ in $k$. Then from (\ref{deformation})
we have
\[
\alpha_1^{(n)}(k)p=k^{n+m+\frac{1}{2}}+O\left(k^{m-\frac{1}{2}}\right),
\]
from which we have
\[
\alpha_1^{(n)}=\left[\frac{k^{n+m+\frac{1}{2}}}{p}\right]_+.
\]
The compatibility between the flows in the hierarchy can be shown 
directly\cite{kodama:02}.
The explicit form of the ${\rm BH}_m$ hierarchy is given by the
following form with the polynomial $\alpha_1^{(n)}(k)$ for $k$ being 
replaced by a
matrix $K$,
\[
\frac{\partial U}{\partial t_n}=\alpha_1^{(n)}(K)\frac{\partial U}{\partial x},
\quad {\rm with}\quad U=(u_0,u_1,\cdots,u_{2m})^T,
\]
where $K$ is the (companion) matrix given by
\[
K=\left(
\begin{array}{llllll}
-u_0 & 1  & \cdots & \cdots & 0\\
-u_1 & 0  & \cdots & \cdots & 0\\
~~\vdots& \vdots& \ddots &\ddots & \vdots\\
-u_{2m-1}&0 &\cdots &\cdots & 1\\
-u_{2m}& 0 & \cdots & \cdots& 0
\end{array}
\right).
\]
For example, $\alpha_1^{(1)}(K)=K-(1/2)u_0I$ with $I=(2m+1)\times(2m+1)$
identity matrix.

It is also interesting to note that the systems in the hierarchy can be
written in the Riemann invariant form,
\[
\frac{\partial \kappa_i}{\partial t_n}=\alpha_1^{(n)}\frac{\partial
\kappa_i}{\partial x}, \quad {\rm for}\quad i=0,1,2,\cdots,2m,
\]
where the Riemann invariants $\kappa_i$'s are the roots of the polynomial
associated with the curve ${\mathcal C}$, i.e.
\[
p^2=\prod_{j=1}^N(k-\kappa_j).
\]
The $u_j$'s are then given by the elementary symmetric polynomials of 
$\kappa_j$'s,
i.e.
\[
u_j=(-1)^{j+1}\sum_{1\le i_1<\cdots<i_j\le 
N}\kappa_{i_1}\cdots\kappa_{i_j}, \quad {\rm for}\quad
j=0,\cdots,2m.
\]
The $\kappa_j$ are also the eigenvalues
of the matrix $K$. Then the solution can be obatined by the generalized
hodograph form\cite{kodama:02},
\[
\Omega_j(v_1,\cdots,v_{2m+1},x,t_1,\cdots)=\sum_{n=0}^{\infty}
\alpha_1^{(n)}(\kappa_j)t_n=0,\quad j=1,2,\cdots,2m+1.
\]
  The regularity of the solution is obtained by the smoothness
of the curve, which corresponds to the distinction of the roots 
$\kappa_j$. The singular structure of the solution can be described by
their intersections of the functions $\Omega_j$ in the hodograph 
solution\cite{kodama:02}.

\subsection{The dispersionless ${\rm JM}_m$ hierarchy.}
The dispersionless ${\rm JM}_m$ (${\rm dJM}_m$) hierarchy is given by 
a dispersionless
(classical) limit of the hidden integrable hierachy of the Jaulent-Miodic
equation\cite{konopelchenko:99}.
The ${\rm dJM}_m$ hierarchy is given by the similar form as the ${\rm BH}_m$
hierarchy\cite{kodama:02}:
\begin{proposition}
\label{dJMm}
For the hyperelliptic curve of genus $m$ given by
\[
p^2=k^{2m+2}+\sum_{j=1}^{2m+2}v_jk^{2m+2-j},
\]
we have an integrable hierarchy of deformation of the curve,
\[
\frac{\partial p}{\partial t_n}=\frac{\partial }{\partial x} Q^{(n)},
\quad {\rm with}\quad  Q^{(n)}=\alpha_1^{(n)}p
=\left[\frac{k^{m+n+1}}{p}\right]_+p.
\]
\end{proposition}
For example in the case with $m=0,~n=1$, we have
\[
\left\{
\begin{array}{llll}
&\displaystyle{\frac{\partial v_1}{\partial t_2}=\frac{\partial}{\partial x}
\left(v_2-\frac{3}{4}v_1^2\right)}\\
&{}\\
&\displaystyle{\frac{\partial v_2}{\partial t_2}=-\frac{1}{2}
\frac{\partial}{\partial x}\left(v_1v_2\right)}
\end{array}
\right.
\]
which describes classical shallow water waves.
Then the ${\rm dJM}_m$ hierarchy can be considered as an integrable system
defined on the singular sector of the codimension $m$ in the solution of the
${\rm dJM}_0$ hierarchy. As in the case of the BH hierarchy,
the singular sectors are given by the intersection structure of the
  hodograph solution $\Omega_j,~j=1,2$ on ${\mathbb 
C}^{\infty}=\{(x,t_1,\cdots,)\}$.

\section{Examples of non-hyperelliptic case}
A natural extension of the hyperelliptic case ($N=2$) may be given by
the case with the curve,
\begin{equation}
\label{nonhyperbolic}
p^N=u_N(k)=k^{M}+\sum_{j=1}^{M}v_jk^{M-j},
\end{equation}
where $N,~M$ are positive integers with $N\ge 3$. The genus of the
({\it irreducible projective}) curve ${\mathcal C}\subset {\mathbb 
C}{\mathbb P}^2$
for $N>M$ is given by Max Noether's genus formula
\[
g=\frac{1}{2}\Big(({N}-1)({N}-2)-(N-1)(N-M-1)-{\rm g.c.d.}(N,N-M)+1\Big),
\]
where g.c.d.($A,B$) implies the greatest common divisor of $A$ and $B$.
If $M>N$, the genus is given by the same formula with the exchange $N\leftrightarrow M$.
Also if $M=N$, the curve is smooth, and the genus is
\[
g=\frac{(N-1)(N-2)}{2}.
\]
For the curve (\ref{nonhyperbolic}), we have a similar result
as in the case of hyperelliptic curve:
\begin{proposition}
\label{Ndeformation}
The following integrable hierarchy of hydrodynamic type provides
a deformation of the curve (\ref{nonhyperbolic});
\[
\frac{\partial p}{\partial t_n}=\frac{\partial}{\partial x}Q^{(n)},
\quad {\rm with}\quad Q^{(n)}=\left[k^{\frac{M}{N}+n}\right]_+p.
\]
\end{proposition}
The compatibility can be shown in the same way as the hyperelliptic case
and it can be found in our paper\cite{kodama:02}.

\begin{remark}
\label{Nrem}
One should note that the hierarchy in Proposition \ref{Ndeformation}
is not the same as a dispersionless Lax reduction\cite{kodama:89}, i.e.
\[
k^N=[k^N]_{+p}={\rm polynomial~of~degree~{\it N}~in~}p.
\]
However the Lax reduction is a special case of the curve (\ref{curve}) where all of the $u_j(k)$'s
do not depend on $k$ except $u_N(k)=-k^N$.
\end{remark}

\section*{Acknowledgments}
YK would like to thank the organizers for a financial support for
this meeting.


\begin{thebibliography}{0}
\bibitem{aoyama:96}
S. Aoyama and Y. Kodama,
Topological Landau-Ginzburg theory with a rational potential and the 
dispersionless
KP theory, {\it Commun. Math. Phys.}{\bf 182}, 185 (1996).

\bibitem{kodama:89}
Y. Kodama and J. Gibbons,
A method for solving the dispersionless KP hierarchy and its exact 
solutions II,
{\it Phys. Lett.} {\bf 135A}, 167 (1989).

\bibitem{kodama:90}
Y. Kodama, Solutions of the dispersionless Toda equation, {\it Phys. Lett.}
{\bf 147A}, 477 (1990).

\bibitem{kodama:02}
Y. Kodama and B. G. Konopelchenko,
Singular sector of the Burgers-Hopf hierarchy and deformations of
hyperelliptic curves, {\it J. Phys. A:Math. Gen.} {\bf 35}, L489 (2002).

\bibitem{konopelchenko:99}
B. G. Konopelchenko, L. Martinez Alonso and E. Medina,
Hidden integrable hierarchies of AKNS type, {\it J. Phys. A:Math. 
Gen.} {\bf 32},
3621 (1999).

\bibitem{zakharov:80}
V. E. Zakharov,
Benny equations and quasi-classical approximation in the inverse
problem method, {\it Funkts. Anal. Pril.} {\bf 14}, 15 (1980).


\end{thebibliography}
\end{document}